# Phase-Modulus Relations in Cyclic Wave Functions

R. Englman*, A. Yahalom# and M. Baer*

*Department of Physics and Applied Mathematics,
Soreq NRC, Yavne 81800, Israel
#Faculty of Engineering. Tel Aviv University,
Ramat Aviv, Israel
(email:englman@vms.huji.ac.il)
Fax: 972 2 6430546

Abstract.

We derive reciprocal integral relations between phases and amplitude moduli for a class of wave functions that are cyclically varying in time. The relations imply that changes of a certain kind (e.g. not arising from the dynamic phase) obligate changes in the other. Numerical results indicate the approximate validity of the relationships for arbitrarily (non-cyclically) varying states in the adiabatic (slowly changing) limit.

1. Introduction

In textbook expositions of Quantum Mechanics the squared moduli of the wave function components , which serve as measures of the state occupation probability, play a central role. In the last 15 years increasing emphasis is put on another quantity, namely on the phases in the wave-function, and in particular on the Berry-phase part which reflects the geometry or topology in which the physical system exists [1-6] and whose practical manifestation is in some interference experiments [7]. By considering cyclic wave functions we derive theoretically and demonstrate computationally connections between phases and amplitude moduli. The connections, shown in Equations (11) and (12) below, take the shape of integral relations and are reciprocal between the two quantities.

2. Formalism.

We first establish reciprocity relations for some periodic functions and then extend the validity of the relations for further classes of functions, including several "cyclic" wave functions and others that are only nearly cyclic.
To start we assume that our function $\phi(t)$ is periodic in t , $\phi(t+2\pi)= \phi(t)$, is suitably smooth, satisfies the relation $\phi^*(t)= \phi(-t)$ for real t (where the star denotes the complex conjugate) and (as a sufficient condition, of which later) that the zeros of $\phi(t)$ lie at points such that Im $t \leq 0$.
 Also, in the trigonometrical polynomial

$$\phi(t) = \Sigma_n a_n \cos(nt) + i\Sigma_n b_n \sin(nt) \tag{1}$$

n goes from 0 to N, which is finite. By dint of $\phi^*(t)= \phi(-t)$ the coefficients are real. Then we define

$$\chi(t) = e^{iNt}\phi(t) = \sum_{m=0}^{2N} c_m e^{imt} \tag{2}$$

The coefficients are related to those in (1), through

$$c_m = (a_{N-m} - b_{N-m})/2 \quad \text{for } 0 \leq m \leq N$$

$$c_m = (a_{N-m} + b_{N-m})/2] \quad \text{for } N < m \leq 2N \tag{3}$$

Therefore also, provided $c_0$ is not zero,

$$\log (\chi/c_0) = \log [1 + (\sum_{m>0} c_m e^{imt}/c_0)] \tag{4}$$

If we now expand $\log (\chi/c_0)$ as (infinite) cos and sine-series

$$\log (\chi/c_0) = \Sigma_n A_n \cos(nt) + i\Sigma_n B_n \sin(nt) \tag{5}$$

since the m's in (4) are positive. We obtain by Abel's theorem [8] that $A_n = B_n$ (all real, since $c_m$ are such). The convergence of the series in (5) is assured by the conditions laid out for $\phi(t)$. Since

$$\log (\chi/c_0) = \log |(\chi/c_0)| + i \arg (\chi/c_0) \tag{6}$$

both functions being real, their Fourier coefficients can be respectively identified with the real coeffficients $A_n$ and $B_n$, whose equality then requires that

$$\int_{-\pi}^{\pi} \log |(\chi/c_0)| \cos(nt)\, dt = \int_{-\pi}^{\pi} \arg (\chi/c_0) \sin(nt)\, dt \tag{7}$$

Now the functions $\cos(nt)$ and $\sin(nt)$ are conjugate pairs that satisfy for all n the following reciprocal relations

$$P\int_{-\infty}^{\infty} dt'\, \cos nt'/(t'-t) = -\pi \sin nt \tag{8}$$

$$P\int_{-\infty}^{\infty} dt'\, \sin nt'/(t'-t) = \pi \cos nt \tag{9}$$

where P represents the principal part of a singular integral.

(To prove these relations, apply to the function $\exp(int)$ Cauchy's theorem with a contour C that consists of an infinite semicircle in the upper half of the complex t' plane traversed anti-clockwise and a line along the real t'-axis.from $-\infty$ to $\infty$ in which the point t'=t is avoided with a small semicircle. Then

$$\int_C dt'\,[\exp(int')]/(t'-t) = 2\pi i\, \exp(int) \text{ or zero} \tag{10}$$

depending on whether the small semicircle is below or above the point t'=t.. The relations in (9) are then got by separating real and imaginary parts [9]). Therefore also the real and imaginary parts of $\log (\chi/c_0)$ satisfy the reciprocal relations, that form the core of this work:

$$P\int_{-\infty}^{\infty} dt'\, \log |(\chi(t')/c_0)|/(t'-t) = \pi \arg (\chi(t)/c_0) \tag{11}$$

and

$$P\int_{-\infty}^{\infty} dt'\, \arg (\chi(t')/c_0)/(t'-t) = -\pi \log |(\chi(t)/c_0)| \tag{12}$$

here proven for functions $\chi(t)$ that satisfy $\chi(t) = e^{iNt}\phi(t)$, where $\phi(t)$ is periodic and has no zeros in the upper half of the complex t-plane.
We shall apply the theorem to the amplitudes in a wave function $\Psi(t)$, which is expanded in some (time independent) set in the form

$$\Psi(t) = \Sigma_i \phi_i(t) |i\rangle \tag{13}$$

Then $\phi(t)$ in (1) may stand for any of the amplitudes $\phi_i(t)$ in the expansion. Before we make applications to "cyclic" or to "adiabatically periodic" wave functions, of the type discussed in [1-6], we note that the physical $\phi_i(t)$'s are not strictly cyclic, since when continuity of the phase is imposed on the wave function, then the phase undergoes changes between periods ( generally in a monotonic manner). We can correct for these phase changes, which include the dynamic phase [2], by multiplying the true wave-function with a phase factor $\exp[i\lambda(t)]$ (with $\lambda(t)$ real) which then leaves us with a properly periodic function. As will be seen shortly, the reciprocal relations provide information on the intraperiod phase variation. The disentangling of the dynamic phase from the total phase observed (say) in an experimental situation will not be treated here, except to say that the former phase is common to all wave-function components in (13), whereas the phases in (11) and (12) are specific to each component (i.e. different for each amplitude $\phi_i(t)$ in (13)].

3. Applications.

As illustration, we consider the time development of a doublet subject to the Schrodinger Equation (in which h=1)

$$i d\Psi(t)/dt = H(t)\Psi(t) \tag{14}$$

whose Hamiltonian in the doublet representation is

$$H(t) = \frac{G}{2} \begin{pmatrix} -\cos\omega t & \sin\omega t \\ \sin\omega t & \cos\omega t \end{pmatrix} \tag{15}$$

Here $\omega$ is the angular frequency of the system, imposed (say) by an external disturbance. The eigenvalues of (15) are $G/2$ and $-G/2$ and we assume that $G>0$. In the former eigenvalue state the amplitude $\phi_1$ shown in (13) is related to a function $\phi'$, in which the dynamic phase factor $e^{-iGt/2}$ has been removed, as follows.

$$\phi' = e^{iGt/2}\phi_1 = e^{iGt/2}\{\cos(Kt)\cos(\omega t/2) + (\omega/2K)\sin(Kt)\sin(\omega t/2)$$
$$-i(G/2K)\sin(Kt)\cos(\omega t/2)\} \tag{16}$$

(cf. 4]) , with

$$K = 0.5\sqrt{(G^2+\omega^2)} \tag{17}$$

For a general K the function in (16) is not periodic, but it will be be such for some choice of K (actually of G/ω which is the basic parameter in the Hamiltonian). Thus we shall choose K/ω to be an integer; then the periodicity of φ in (16) is 4π/ω (compared to 2π in the "mathematical" section). We note that in terms of the parameters appearing in the above equations the adiabatic (AD) limit, which is the subject of the subsection (B), is characterised by

$$K/\omega \approx G/\omega >> 1 \qquad (AD) \tag{18}$$

(A) Fast, non-adiabatic motions

The cases that will be now be presented are where K/ω is an integer, not necessarily large. Examining the function φ in (16), we note that the quantity corresponding to N in (2) and connected to the fastest oscillating term is given by

$$N = (2K+\omega)/(\omega) \tag{19}$$

which is an integer. Under these conditions φ is cyclic and also satisfies the other conditions in Section 1 (see below). We can thus test the reciprocal relations (11) and (12) on it. The result is shown graphically in figure 1 for K/ω =1.

------------
Fig.1
-----------

The curves computed directly from (16) (full lines) coincide so accurately with those obtained from the integral relations (broken lines) that the results need to be exhibited as mirror images. The same occurs for K/ω being any integer different from one. The phases of φ' [given in (16)] and χ [shown in (2)] differ by (G -ωN) t / 2 and the phase of the former (which is physically significant) is shown by the curves.
     It may be thought that we ought to work directly with the physical phase of $\phi_1$. However the insertion of this phase in the reciprocal relation will lead to a problem, since the infinite-range integral in (12) diverges for a phase that is a linear function of t. The avoidance of these divergences by a proper choice of the function and having a phase that is of bounded variation are the basic features of the formalism. They allow one to obtain a unique phase for χ.
     Note also that our results provide what is called the "connection", namely the behaviour of the phase within a period. The integral of the "connection" calculated over a "full" revolution is essentially the "Berry

phase" [1,3]. We obtain this phase by first calculating the phase of $\chi$ and then amending it by the previously mentioned phase difference. For our model this phase is then given by

$$2\pi + (G - \omega N)\pi/\omega = [1 - (2K-G)/\omega]\pi < \pi = \text{the adiabatic limit} \tag{20}$$

The curves in the figures show agreement with this value.

Note, that in order that $\chi$ be periodic, its phase must make a jump at the edges of the period. (This jump has no physical significance, in contrast to the smooth variation of the connection between the extremities. It is not shown in fig. 1a or in the following figures.) Put differently, the derivative of the phase at the extremities is not smooth, but shows a delta function behaviour. It can be shown, that this implies that the log of the amplitude modulus must also be singular at these points, or that the amplitude has a zero there. This is indeed the case for $\phi'$ given in (16), from which equation it can be checked that

$$\phi'(\pm \pi/\omega) = 0 \tag{21}$$

All further zeros of $\phi'$ or $\phi_1$ lie at points such that Im $t < 0$, true for all integer values of $K/\omega$. In other cases this may not be satisfied and the assumptions of the theory are then violated, requiring it to be generalised [11].

We have also confirmed, by numerically integrating the two members of Eq. (7), that the equalities for the Fourier coefficients $A_n$ and $B_n$ in (4) are satisfied, at least up to n=50. The asymptotics of these coefficients for high n clearly exhibit the singularities mentioned in the previous paragraph. Our numerical work also indicates that the expansion of (4) does not converge uniformly, but is summable in some mean sense so that, by Fejer's theorem ([10], p. 254), the Fourier series converges to the (finite) value of the logarithm in (4)..

(B) Adiabatic and nearly adiabatic cases.

The adiabatic limit was treated by Berry [1] and in numerous publications since, e.g. [3]. The "connection" exhibits a step function behaviour and the log of the amplitude a delta-function-like dip. In the near adiabatic case, as one approaches $t = \pm \pi/\omega$, one obtains, in the phase and to a lesser extent in the amplitude modulus, a series of oscillations with period of $2\pi/K (<< 2\pi/\omega)$. One can study the behaviour of the phase in the vicinity of these points from expression (16) for $\phi$. Keeping only first order terms in $(t - \pi/\omega)$, we find that the phase near $t = \pi/\omega$ can be approximated by

$$\text{Im ln}\,[(t - \pi/\omega)) + \sin(Kt)\exp(iKt)/K] \tag{22}$$

In figure 2 we compare the directly calculated curves with those derived from the integral relation. The comparison and confirmation of the reciprocal relations are made for $K/\omega=17$ which departs significantly from the adiabatic limit ($\omega\to 0$), but in that limit the oscillations would not be visible. In any case, the agreement gets better the more adiabatic the system.

-----------
Fig.2
-----------

One would (of course) expect similar agreement for the neighbouring periodic case, when $K/\omega=16$, and indeed for any integral choice for this parameter. The question is now, what happens for a general value of this parameter (say, between 16 and 17) in this, the adiabatic limit. Figure 3 has been computed for $K/\omega =16.59$ (for which the solution is not periodic). Though the reciprocal relations (based on periodic solutions) are not expected to hold exactly in this case, the similarity is remarkable, especially as regards the positions of the peaks. They are probably sufficient for experimental purposes (e.g. for comparison between observations made directly on the phase and those that derive it from state occupation probabilities). The extra, outsize peaks in the curve got from the integral are, at least partly, manifestations of the Gibbs phenomenon in Fourier series [8,10] and are irrelevant to Physics.

--------------
Fig.3
--------------

We may thus tentatively conclude that close to the adiabatic limit the reciprocal relations (11) and (12) between absolute amplitudes and phases will generally provide a good approximation. In a future publication we shall generalise the theory to cover other cyclic and noncyclic situations.[11]

4. Discussion

    The import of the theory developed in this paper is manifold. We point to wave-packet reassembly ([12] ), to polarisation currents in extended systems (moving in a periodic k-space)[13,14], to the theory of quantum measurements [15]. Some of these are under our consideration. As a curiosity, we mention a text-book relevance of reciprocity relations, that hold for the oscillating wave packet in a harmonic potential.[16].

    Previous connections between phase and modulus amplitude are implied in the equation of continuity, for the special case of coordinate-space representation, and in a heuristically proposed relation for polarisation in solids ([14] and references therein.). The relations in (8) and (9) are formally identical to the well-known Kramers-Kronig relations (between real and imaginary parts of response functions), which operate in the frequency plane. The origin of the latter is in the principle of causality, leading to a real-time asymmetry (due to the imposition of initial conditions). In contrast, the relations found in this

Letter arise from the structure of the Schrodinger equation which contains an imaginary-time asymmetry.

Derivatives of $\arg\chi$ and of $\log|\chi|$ feature in tunneling situations and are in fact the components of the Buttiker tunneling time. The forms given in ,e.g., [17] for the real and imaginary delay times satisfy quite well reciprocal relations in the frequency domain.

At this stage the following appear to be the theoretical significance of the reciprocal relations :

(a) They show that changes (of a nontrivial type) in the phase imply necessarily a change in the occupation number of the state components and vice versa.

(b) One can define a phase that is given as an integral over the log of the amplitude modulus and therefore is an observable and is gauge-invariant.

(c) Experimentally, phases can be obtained by measurements of occupation probabilities of states using (11). (Cf. [15].)

(d) Conversely, the implication of (12) is, as noted by a referee, that a geometrical phase appearing on the left-hand-side entails a corresponding geometrical probability, as shown on the right-hand-side. Geometrical probabilities have been predicted in [18] and experimentally tested in [19].

Figures.

1. Test of the reciprocal relations from comparing inputs and outputs of Eqs. (11) and (12) for the state amplitude $\phi(t)$ in (16)
The non adiabatic, cyclic case ($K=\omega$, $G=\omega\sqrt{3}$.)
(Upper). The phase angle against time for the positive half of the period. The phase is anti-symmetric upon time inversion..
The solid curve is from direct computation of arg $[\phi(t)]$ and the broken curve is obtained from the numerical integration of (12), into which the computed values of the amplitude modulus are fed. The y-axis has the same sign upwards and downwards, since otherwise the curves would be indistinguishable. Analogously to the Berry phase, near the boundaries there is a net change of $0.75\pi = 2.36$, which is in agreement with the result in (20) for this non adiabatic case.
(Lower) The log of the amplitude modulus against time, decreasing for both the input (full lines) and output (broken lines). The modulus has time inversion symmetry.

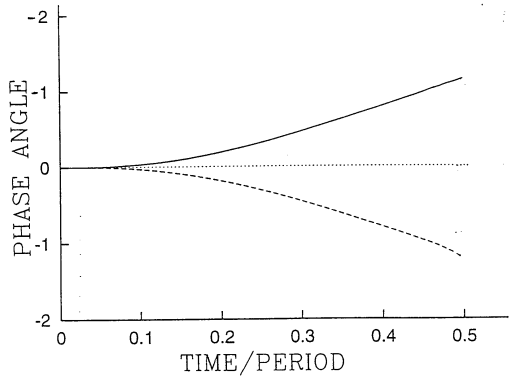

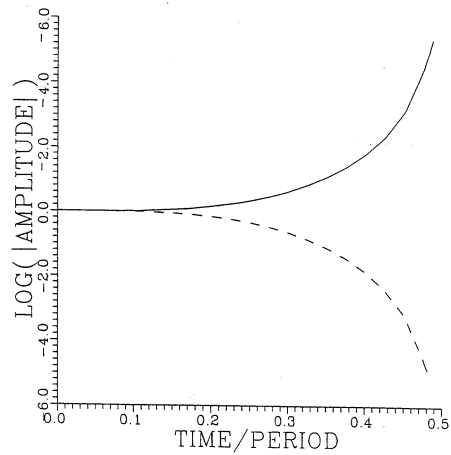

Fig. 2. Comparison when the theory predicts agreement. A near-adiabatic and cyclic case (K=17ω, G=ω√1155). Full lines =input, broken lines= output.
(Upper) The phase.
 Note the appearance of half the Berry-phase at the positive edge, the steep rise there and the undulations, akin to Rabi oscillations.
(Lower) The modulus.
 The oscillations are barely visible on the logarithmic plot, but are evident in a linear plot of the modulus.

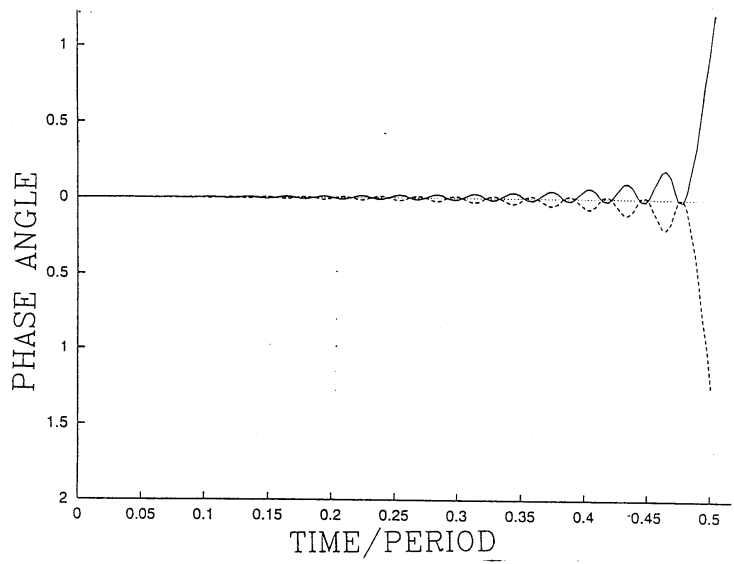

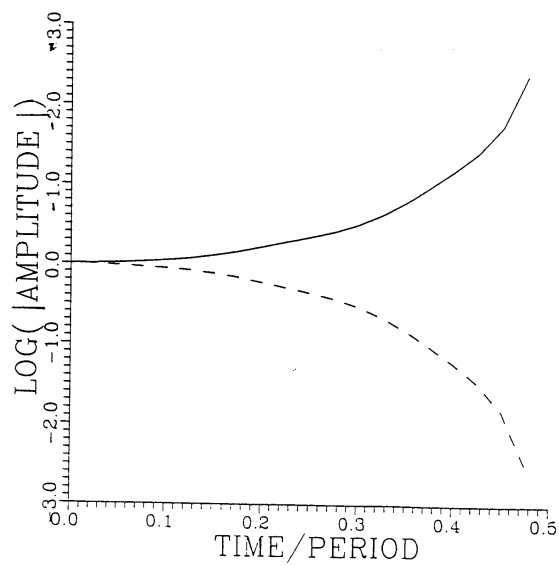

Fig. 3. A "general", near adiabatic and non-cyclic case.
(K=16.59ω, G=ω√1100). In this case the assumptions of the theory do not hold, still there is a fair agreement between input (full curve) and output (broken curve) (Upper). The phases are plotted with their true signs. One sees that there is a slight vertical offset between input and output, but the oscillations are in the same place. For better appreciation, the phases are shown for double the range in the other figures. (For the sharp peaks, see text.)
(Lower). The modulus against time for half a period.
Marked undulations are apparent in the two curves, which are practically indistinguishable.

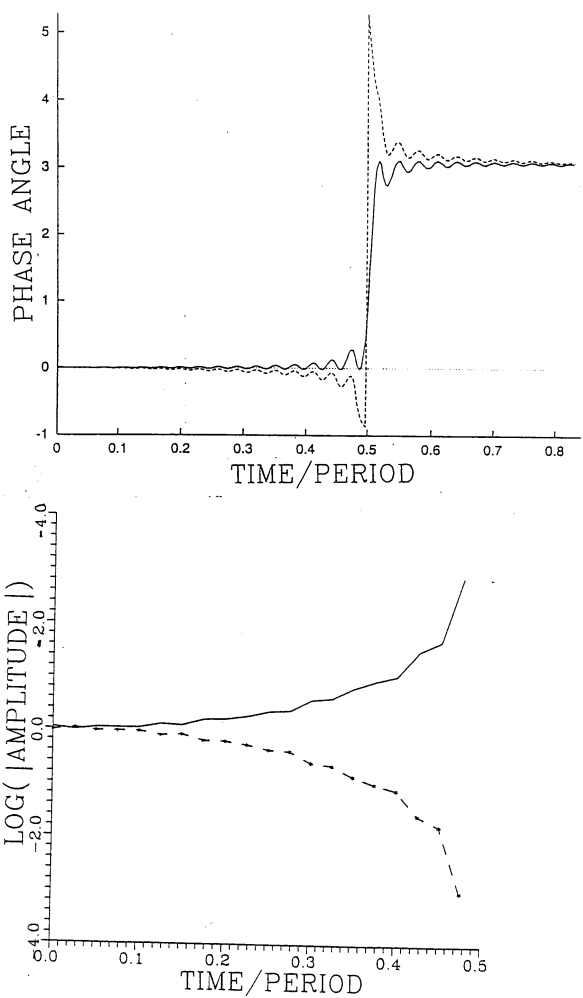

Fig. 3